\RequirePackage{fix-cm}
\documentclass[smallextended]{svjour3}       
\smartqed  
\usepackage{float}
\usepackage{graphicx}
\usepackage{amsfonts}
\usepackage{grffile}

\begin{document}

\title{Sanction or Financial Crisis?
}
\subtitle{An Artificial Neural Network-Based Approach to model the impact of oil price volatility on Stock and industry indices}


\author{Somayeh Kokabisaghi \and Mohammadesmaeil Ezazi\and Reza Tehrani \and Nourmohammad Yaghoubi  
}

\institute{1.Somayeh kokabisaghi \at
              Faculty of management and economics, University of Sistan and Baluchestan, Zahedan, Iran \\
              Centrum Wiskunde \& Informatica, Amsterdam, Netherlands\\      \email{karen.kokabisaghi@cwi.nl}, ORCID: 0000-0002-4589-7638          
           \and
           2. Mohammadesmaeil Ezazi \at
              Faculty of management and economics, University of Sistan and Baluchestan, Zahedan, Iran\\ \email{mohammad.e.ezazi@gmail.com}
              \and
           3. Reza Tehrani \at
              Faculty of Management, University of Tehran, Tehran, Iran\\ \email{rtehrani@ut.ac.ir}
              \and
           4. Nourmohammad Yaghoubi \at  Faculty of management and economics, University of Sistan and Baluchestan, Zahedan, Iran\\ \email{nm.yaghoubi@gmail.com}
}

\date{Received: date / Accepted: date}

\maketitle

\begin{abstract}
 In this paper, we model the impact of oil price volatility on Tehran stock and industry indices in two periods of international sanctions and post-sanction. To analyse the purpose of study, we use Feed-forward neural networks. The period of study is from 2008 to 2018 that is split in two periods during international energy sanction and post-sanction. The results show that Feed-forward neural networks perform well in predicting stock market and industry, which means oil price volatility has a significant impact on stock and industry market indices. During post-sanction and global financial crisis, the model performs better in predicting industry index. Additionally, oil price-stock market index prediction performs better in the period of international sanctions. Herein, these results are, up to some extent, important for financial market analysts and policy makers to understand which factors and when influence the financial market, especially in an oil-dependent country such as Iran with uncertainty in the international politics.


\end{abstract}

\section{Introduction}
\label{intro}
Crude oil price fluctuations have been a concern for the world macroeconomy since oil crises in 1970, 2008 and 2014. According to Organization of Petroleum Exporting Countries (OPEC), the oil price dropped from \$145 to \$30 in mid 2008 and reached to the low price of \$27 in 2014. These sharp downward trends (shocks) influence the economy by disturbing aggregate economic activities and spread to stock market and energy indices \cite{Guntner2014} and \cite{Nusair2016} and \cite{Zhang2016} and \cite{Bastianin2016} and \cite{Wen2012}and  and \cite{Angelidis2015} and \cite{Niknam2016} and \cite{MoyaMartinez2014}.
Some studies reported that oil price shocks have different effect in different economies. For example, oil price shocks have different effect on the U.S. economy and oil-exporting countries \cite{Kilian2009} and \cite{Wang2013}. 
Although it is expected that higher oil price leads to higher revenue, cash flow and therefore growth in the economy and financial markets in oil-exporting countries \cite{Arouri2010} the exact impact of the oil price changes on the financial markets is still unclear. 
Furthermore, different factors such as the source of oil price shocks \cite{Kilian2009}, political issues, developed or emerging stock markets and whether the country is oil-exporter or oil-importer \cite{Wei2017} have made it more difficult to draw a clear conclusion on the effect of oil price shocks on financial markets \cite{Basher2006} and \cite{Wang2013}. Hence, understanding the underlying behaviour of oil price is important to keep track of changes in the target economy. 

On the other hand, Oil price shocks and its contagion on other economic indices and prices have made modelling and prediction complicated \cite{Kokabisaghi2018}. Earlier studies used econometric models such as vector autoregressive (VAR), generalized autoregressive conditional heteroskedasticity (GARCH) \cite{Park2008} and \cite{wei2010}. But the complexity and nonlinear behaviour of oil price and financial and economic variables have convinced researchers to use artificial intelligence methodologies to deal with unpredictable changes in oil price and other economic variables \cite{mingming2012} and  \cite{Gurusen2011} and  \cite{Bissoondeeal2011}. Overall, several researches have been devoted to analyze the oil price shocks, its origins and impacts on economic factors and financial markets. However, there is surprisingly a lack of focus on oil-dependent economies such as Iran.

Since Iran is one of the largest oil exporting countries, both oil crises and political tensions can impact its economy. During global financial crisis in 2008, Iran economic growth decreased to the lowest rate 1.8 percent \cite{WorldBank2018}. Despite the fact that high oil price is beneficial for oil-exporting countries \cite{Korhonen2010} and lower oil price creates instability in oil dependent countries \cite{Kitous2016}, it was expected that Iran economic growth increases after the global financial crisis and rising oil price; but Iran economic growth had a downward rate of -0.2 percent in 2013. One possible reason is international sanctions imposed on Iran’s industries and banking system. Sanctions which targeted oil created many restrictions in exporting oil and foreign investments in energy industry.
Being largely oil dependent, Iran oil exports dropped from 2231.980 barrel/day to 1081.145 barrel/day during 2009 to 2015. Alongside that, foreign investment was decreased from \$ 3773.8 million to \$ 945 million. 
Generally speaking, it is expected that Iran stock market is affected by uncertainty in international oil market and political tensions. While \cite{oskooe2012} reports that there is no evidence on the impact of oil price volatility and Iran stock market. Differently, \cite{Salehi2015} finds that there is strong causality between oil price volatility and stock price in Iran. Besides, changes in macroeconomic variables affect stock market.

An overview of Tehran stock exchange shows that Tehran Stock Exchange Price Index was increased from 25035.2 million  to 78849.3 million Unit during heavy sanctions (from 2012 to 2014) and declined to 61426.1 million unit in global financial crisis to 2015 and increased to 99414.5 in 2018 \cite{tes2019}. The growth in Tehran stock exchange during sever sanctions maybe explained by \cite{Biglaiser2020} that finds that sanctions impact stock market in the targeted country negatively and significantly only if targeted country were not already subject to multiple sanctions. 
So far, no studies have compared the effect of financial crisis and comprehensive sanctions together on Iran stock and industry indices.

In this paper we aim to model the impact of oil price volatility on stock market and industry indices in Iran. In particular, we investigate how potential uncertainty in oil price and Iran energy industry caused by political tensions and economic crises influence financial and industry indices. 

The reason why Iran is an ideal case for the purpose of this study is because Iran is a member of OPEC and one of the largest oil- exporting countries that can influence the supply side of the international oil market. Iran economy has been under severe international sanctions and witnessed several oil crises while more than 60 percent of Iran revenue is within the oil market \cite{Farzanegan2009}.

Therefore, knowing the impact of international sanctions and/or financial crises on Iran economy and whether international sanctions have been successful to meet their target, is a game changer for Iran policy makers and international politics. Moreover, the source of volatility in the stock market and industry is useful from trading and practical perspective. 
 
To shed light onto the aim of the paper, we take an inspiration of several studies that proved the accuracy of the artificial intelligence as a methodology to model the effect of unusual behaviours in oil price on stock and industry indices \cite{Ince2019} \cite{Atsalakis2009} and \cite{Onder2013} and \cite{Svitlana2016}. Additionally, we compare oil-stock nexus in two periods of international sanctions and post sanctions. 

The reminder of the paper is organized as follow. Section 2 is an overview of financial crisis, international sanctions and methodology, section 3 presents data and architecture of the model, section 4 is the empirical results and section 5 is the conclusion. 

\section{Literature Review}
\label{sec:literature}
\subsection{Overview of Oil price shocks and International Energy Sanction}
\label{sec:event_summary}

Sanction as a pressure tool has been used by policy makers to make changes in nation's policies or achieve certain objectives. Generally speaking, sanctions have a direct impact on the Achilles heel of the target. 

In 2007, United Nation Security Council imposed sanctions on Iran to enforce this country to suspend nuclear activities and also meet the requirements of IAEA (United Nations Security Council. Sanction Resolution no. 1747: UN; 2007. Security Council of United Nations. Resolution no 1929; 2010) It continued till 2010 and banned Iran from any activities related to ballistic missiles and blacklisted all entities and individuals involved with this program such as travelling and financial services.

In the case of Iran, international sanctions have been imposed on the energy sector and banking system.
Both economic and energy sanctions have put a sever strain on oil exports and developments that the offshore supergiant South Pars natural gas fields needed \cite{Sabatini2010}. In November 2011, The US, UK and Canada imposed bilateral restrictions on Iran’s oil and petrochemical industries; UK enforced all British financial institutions to stop doing business with Iranian counterparts. Furthermore, US threatened all countries for having any deal with Iran. Although investment on Iran energy industry was beneficial for European countries, but they kept their strategies against Iran. Otherwise, EU had to deal with the risk of losing international trades with US.
In 2012, the European Union banned importing crude oil and petroleum products from Iran. Before European Union sanctions (EU), Iran oil export was around 2.2 million barrel/day. But in 2012 when EU sanctions came into effect, Iran oil export dropped to 1 million barrel/day. In addition to that, Iran lost non-EU buyers (China, India, Japan, South Korea and Turkey) and oil exports declined to more than 50 percent compared to the past years. Sanctions and poor economic health led to high inflation, unemployment rate and devaluation of the national currency \cite{yong2013}.   

When US tightened sanctions on Iran central bank, Iran was disconnected from the SWIFT (electronic financial transactions). Sanctions on oil trades not only disposed Iran of foreign investment flow, it also impressed Iran’s share in gas sector by disposing access to energy technologies such as LNG technology, which is important for competitiveness in the gas market. As a result, Iran was not able to exploit gas. Furthermore, the national currency, Rial, fell to its lowest value against the US dollar more than 80 percent since 2011. Thus, the government had no choice but to borrow from its Central Bank, which resulted in an increase in the money supply and inflation \cite{Ghorbani2018}. Iran gross domestic product (GDP) growth was deprecated 8.156 \% and -7.445 \% from 2009 to 2012 respectively \cite{WorldBank2018}. Although the sanctions influenced Iran economy by heavy restrictions on oil production and export, but this impact was temporary. Iran changed oil contracts and found new export markets by price concessions. The main Iran oil buyers were China (22 percent), India (13 percent) and Japan (14 percent). In addition, Iran gas export increased from 5.670 billion cubic meters to 9.307 billion cubic meters \cite{IMF2018}.

In 2014, Iran GDP growth increased to 4.603 \% by Optimism in nuclear deal between Iran and the world powers and easing part of sanctions on Iran oil export \cite{WorldBank2018}. Simultaneously, oil price dropped from \$109.62 to \$41.5 in 2014 to 2015. Being largely oil dependent, Iran economic growth declined to - 1.321 \% ultimately \cite{WorldBank2018}. 
Although after global financial crisis and post sanction, Iran export recovered in 2018 to the earlier level before the sanctions 2125.000 b/d but Iran economy suffers from instability, high inflation, drastic devaluation of national currency and stock market inefficiency.


\subsection{Overview of methodology}
\label{sec:model_explenation}

As we previously mentioned, the unpredictable behaviour of financial time series such as crude oil price and stock market index make the analysis difficult. Some studies used econometric models to show the correlation between oil price volatility and stock market; for example, Wei and Guo (2017) applied VAR (vector auto regressive) to show the effect of oil price on stock market in China; (see also, \cite{Kang2015} and \cite{Pandey2018} and \cite{Huang2016}). Some other researchers reported that the real world systems are often nonlinear, thus, it is unreasonable to use linear statistical methods that are built on linear assumptions.
To overcome the linear limitations, researchers have proposed several classes of nonlinear models such as autoregressive conditional heteroscedastic (ARCH) model \cite{Engle1982}, general autoregressive conditional heteroscedastic (GARCH) \cite{Bollerslev1986} among others. However, these models perform well for specific nonlinear patterns and they are not able to analyse other types of nonlinearity in time series. 

To explain the non-linearity of various financial time series, studies used artificial intelligence methodologies (\cite{Lu2011} and \cite{Ticknor2013} and \cite{Kristjanpoller2015} and \cite{Gurusen2011} and  \cite{Bissoondeeal2011}). The results of researches show that artificial neural networks (ANN) is a better method for simulating unanticipated features of financial time series. On reason is because ANN is data- driven and non-parametric. In addition, no prior assumptions of the model form is required and ANN learns from examples to capture the relationships among the data even if the underlying linkage is unknown \cite{Ince2019}.
In addition, ANN with simple architecture can be applied to different situations in finance and economics \cite{Galeshchuk2016} and \cite{Fahima2018}. Furthermore, ANN has ability to capture subtle fractional relationship between variables even in time series with different features such as shocks \cite{Atsalakis2009} and \cite{Onder2013} and \cite{Svitlana2016}. The universal approximation theory also suggests that a single hidden layer neural network can interpret any input-output structure sufficiently \cite{Ince2019}.
\subsubsection{The feed-forward architecture}
\label{sec:model_architecture}
\begin{figure} [hbt!]
  \includegraphics[width=0.75\textwidth]{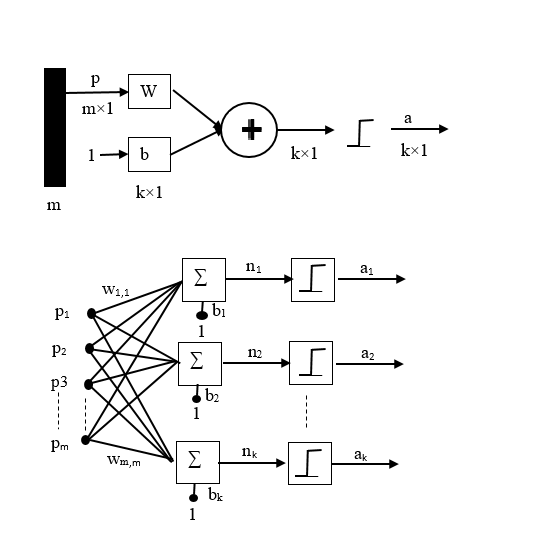}
\caption{This figure represents a Feed-forward neural networks architecture. m is the number of inputs shown by p, k is the number of neuron's layer, W is synaptic weight, b is bias and a is hardlimit function}
\label{fig:schematic_network}       
\end{figure}

\begin{figure} [hbt!]
   \includegraphics[width=0.75\textwidth]{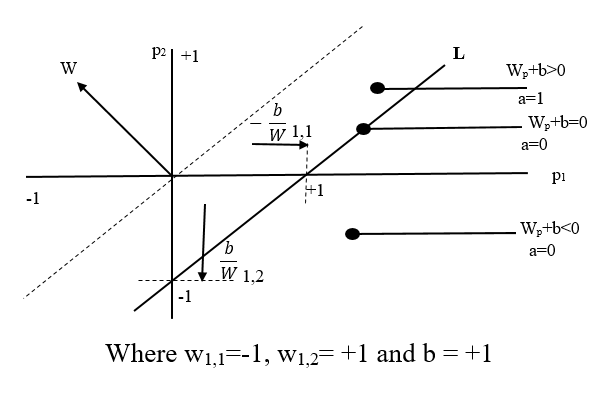}
\caption{This figure represents hardlimit transfer function performance that classify net inputs; if net input to the function reaches a threshold, it forces a neuron to output 1, otherwise it outputs zero }
\label{fig:hardlimit}       
\end{figure}

The feed-forward neural network in this study is a layered network with fully connected hidden layers and outputs. In particular, Feed-forward network can arbitrarily and precisely approximate functions with many finite discontinuities as well as their derivatives. Learning the neural networks is important to optimize the architecture of the network by modifying the weights. If learning is done properly the neural network can update connections of neurons and modify weighted function data.
The main steps for learning networks are first initializing the network weights and comparing the error values between calculated and observed outputs to find the correction vector. Then, the weights for connections between errors are recalculated by determining the correction vector. figure \ref{fig:schematic_network} and \ref{fig:hardlimit} represent a feed-forward neural network and the activation function.

The mathematical structure of the network is shown as follow: 
\begin{equation}
    u_k= \sum^{m}_{j=i} W_{kj}\, p_j 
       \label{neuron_architecture_u}
\end{equation}
where $u_k$ is the output of the adder (sum of the weighed input signals)

\begin{equation}
 \mbox{hardlimit function} \quad \quad \quad \quad
 \\
 a=\mbox{hardlim}(W_p+b)
\end{equation}

\begin{equation}
     y_k= a (u_k+b_k)
         \label{neuron_architecture_y}
\end{equation}
where $y_k$ is the output signals of the neurons
\section{Data and Methodology}
\label{sec:data_methodology} 
\subsection{Data}
\label{sec:data_description} 

To analyze the aim of study, we used information from Iranian Central bank, Tehran Stock Exchange and Organization of the Petroleum Exporting Countries (OPEC). The data comprises daily prices and values for oil, gas and gold price, exchange rate \footnote{The rate of Rial to 1 Dollar}, stock market index (TEPIX) \footnote{Tehran Price Index}, industry index and turn. The empirical study covers 10-year of daily datasets from December 2008 to December 2018 (international energy sanctions from December 2008 to 2014) and (global financial crises and post sanction from 2014 to 2018).

The reason for choosing OPEC oil price is because Iran is a member of OPEC and international crude oil prices follow the same trends more or less. 
Table no. 1 summarizes the descriptive statistics associated to the research variables. In descriptive information in Table \ref{tab:description} shows a heavy tailed distribution for most of time series, which can be explained by the fact that these financial and economic variables witnessed global financial crisis in 2014 and international energy sanctions. 
%
\begin{table}[hbt!]
\caption{Descriptive statistics (Daily data from 2009 to 2018)}
\label{tab:description}       
\begin{tabular}{llllll}
\hline\noalign{\smallskip}
& Mean&Median&SD &Skewness&Kurtosis\\
\noalign{\smallskip}\hline\noalign{\smallskip}
 Oil price (USD)& 77.2 &  75.06 &27.29 &-\,0.03 &-\, 1.4 \\
 Gas price (USD)&3.48&  3.42&0.91&0.56 &0.7\\
 Gold price (USD)& 13.7& 1,275&219.6&0.4&-\,0.2\\
 Exchange rate (Rial)&25,573&31,345&11,657&-\,0.2&-\,1.57\\
 Stock Index (Unit)& 48.779&56,784&28,487&  -\,0.02&-\,1.5\\
 Industry Index (Unit)& 40854.5&48,468&24,791&0.01& -\,1.6\\
 Trading volume (Million)&626.672193&424.083128&893.369438 & 10.19& 183.3\\
 
\noalign{\smallskip}\hline
\end{tabular}
\end{table}

\subsection{Model specification, Feed forwards neural network}
\label{feedforwardNN}

In this paper, we use Feed-forward neural network (FFNN) to analyse the purpose of our study in two periods of International energy sanction and post-sanction. The architecture of our model is the following: 

FFNN is developed with an input layer consisting of five neurons that is five inputs including oil price, gas and gold price, exchange rate, trading volume and the output layer has two neurons that represents dependent variables, stock market and industry indices. In every period of the study, the neurons of the hidden layer are computed as follow:
\begin{equation}
\mbox{Neurons in Hidden Layers}= \frac{1}{2}(\mbox{Inputs}+\mbox{Outputs})+\\
\sqrt{\mbox{nr.training patterns}}
\end{equation}

In order to improve the performance of FFNN, we scale the data between 0,1 as follow:
\begin{equation}
    x_{scaled} = \frac{x - x_{min}}{x_{max} - x_{min}}
\end{equation}

Finally, For FFNN estimation, We split the datasets into two period of sanction (from 2009 to 2014) and post-sanction (2014 to 2018). For each period of study, we use 75 \% of the dataset for training, 20 \% for test and 5\% for validation purpose.
The activation (transfer) function in FFNN is hardlimit to find the relationships between input and output nodes in the network. 
{R1.1}
At the end, we check the estimated RMSE and MAPE to assess the accuracy of networks as follow:
\begin{equation}
RMSE=\sqrt{\frac{1}{N}\sum_{t=1}^{N} (s_t-o_t)^2 }
\end{equation}
\begin{equation}
MAPE=100 \, \frac{1}{N} \,\sum_{t=1}^{N} |\frac{s_t-o_t}{s_t}|
\end{equation}
Where $s_t$ and $o_t$ are actual and predicted values at time t respectively, and N is the number of observed data.

\section{Empirical results}
\label{sec:emprical_result} 

\subsection{Learning Feed-forward neural}
\subsubsection{The period of International energy sanction:}
\label{sec:Analysis_period_sanction} 
The first dataset includes 1845 data from December 2008 to 2014 when severe international energy sanction was tightening on Iran.
There are five inputs including oil, gas and gold price, exchange rate, trading volume as independent variables, and stock market and industry indices as outputs.
As we mentioned in section \ref{feedforwardNN}, the datasets are normalised and split into 75 \%training, 20 \% test and 5 \% validation set and 40 nodes in hidden layers are computed. 

\begin{figure} [htbp]
 \includegraphics[width=0.5\textwidth]{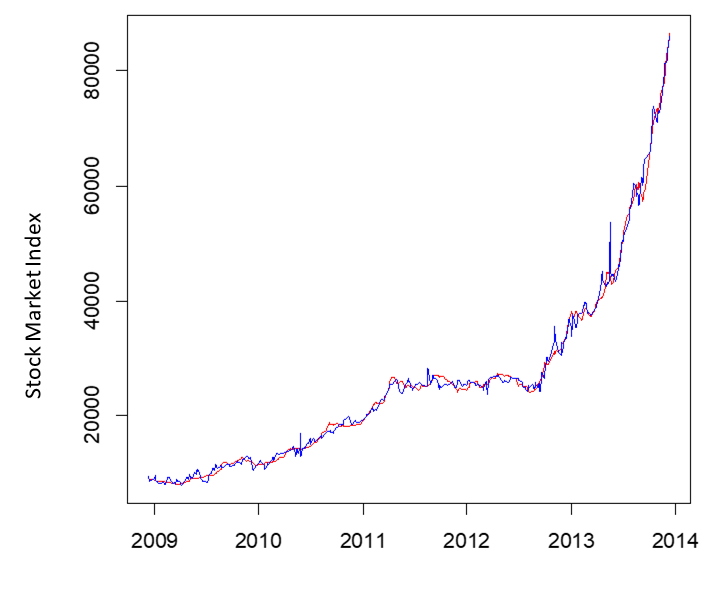}
 \includegraphics[width=0.5\textwidth]{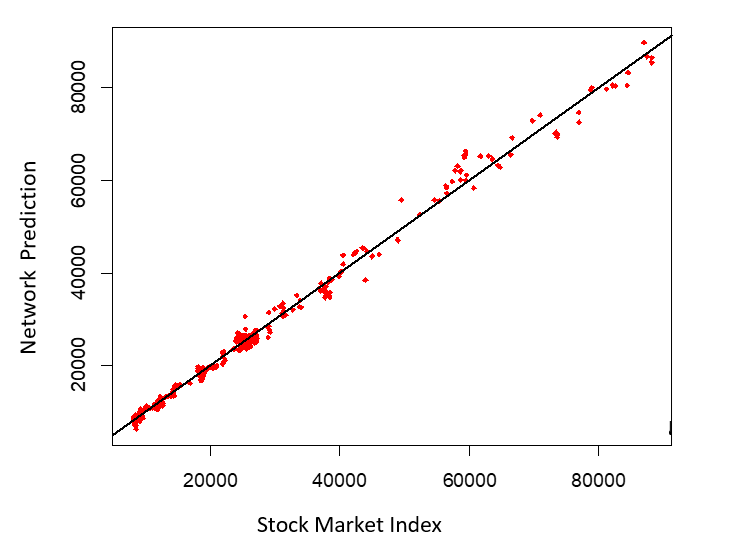}

\caption{The left figure represents TEPIX prediction and actual data; The red line is actual values and blue line is feed-forward network prediction; The right figure shows TEPIX and fitted line (in the period of international energy sanction) }
\label{fig:stock_sanction}       
\end{figure}

\begin{figure} [htbp]

\includegraphics[width=0.5\textwidth]{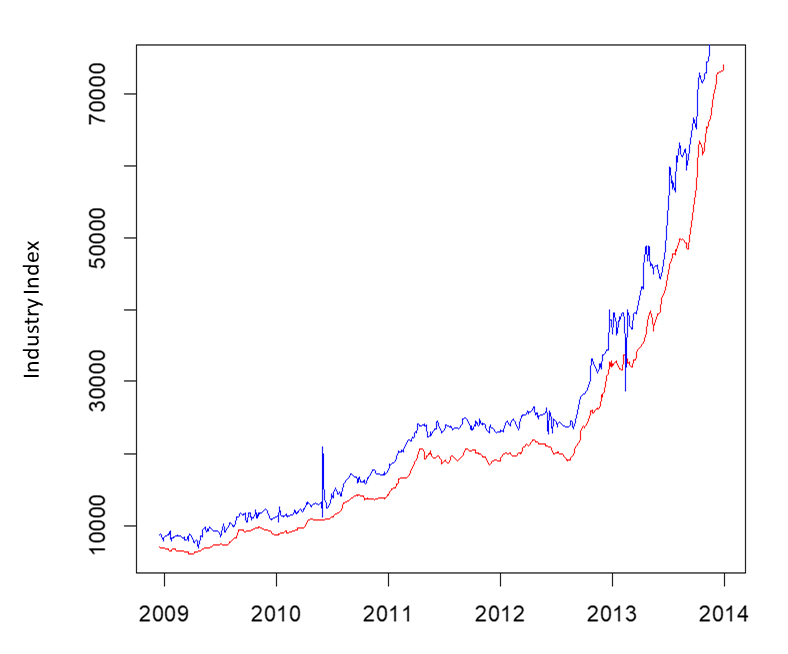}
\includegraphics[width=0.5\textwidth]{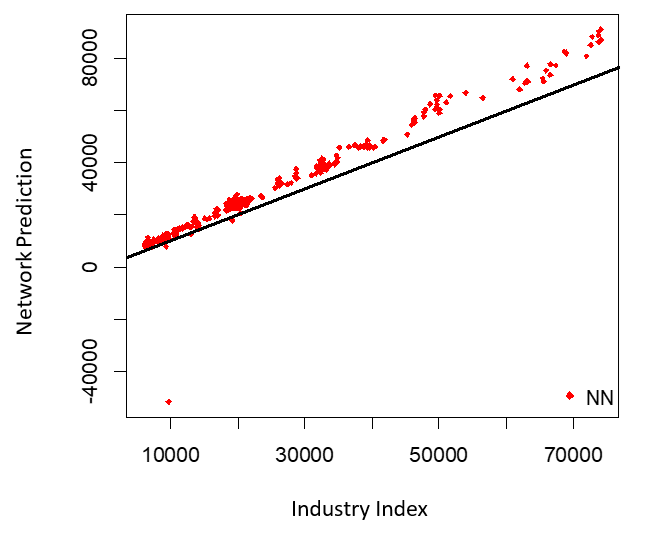}
\caption{The left figure shows Industry index prediction and actual data; The red line is actual values and blue line is feed-forward network prediction; The right figure represent Industry index and fitted line (in the period of international energy sanction) }
\label{fig:industry_sanction}       
\end{figure}

The results of learning feed-forward network presented in Figures \ref{fig:stock_sanction} (right panel) shows the actual datasets is close to fitted line (perfect fit) and there is no significant deviation between the prediction and the actual values.
Figure \ref{fig:industry_sanction}(right panel) represent the perfect fit for industry index approximately. 
Overall, the results of learning network provide 90 percent accuracy for both indices.

\subsubsection{Post sanction and global financial crisis}
\label{sc:post_sanction}
\begin{figure} [hbt!]
 \includegraphics[width=0.5 \textwidth]{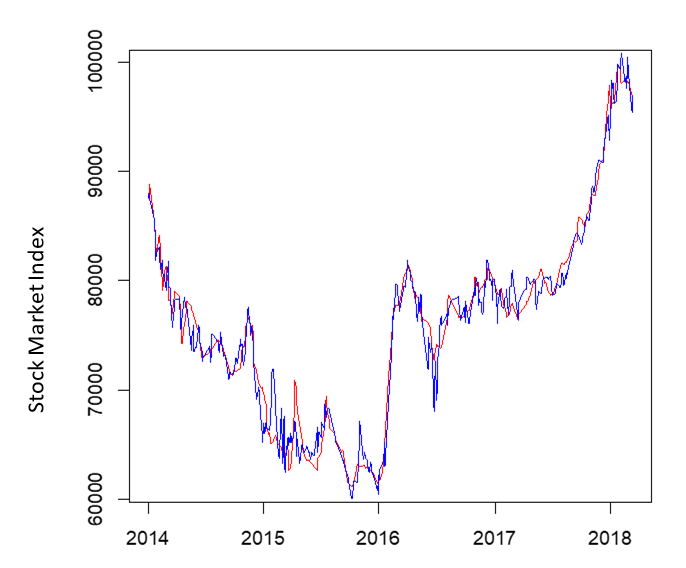}
  \includegraphics[width=0.5 \textwidth]{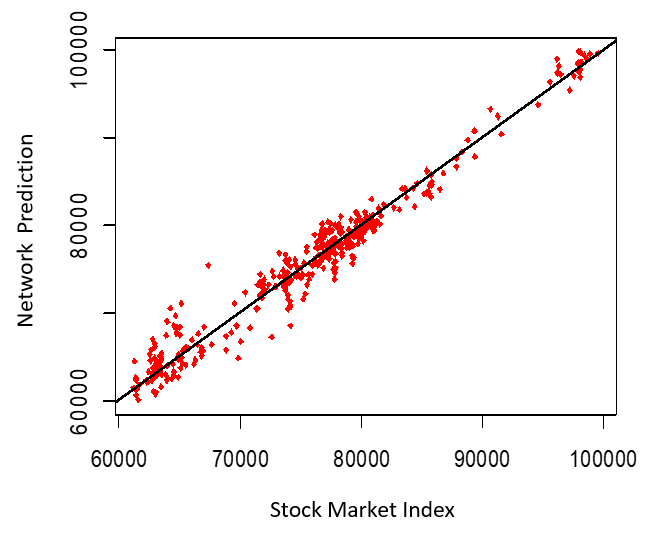}
\caption{The left figure shows stock market index prediction and actual data The red line is actual values and blue line is feed-forward network prediction; The right figure shows stock market index and fitted line (perfect fit) (In post-sanction and global financial crisis) }
\label{fig:stock_post}       
\end{figure}

\begin{figure}[hbt!]
\includegraphics[width=0.5 \textwidth]{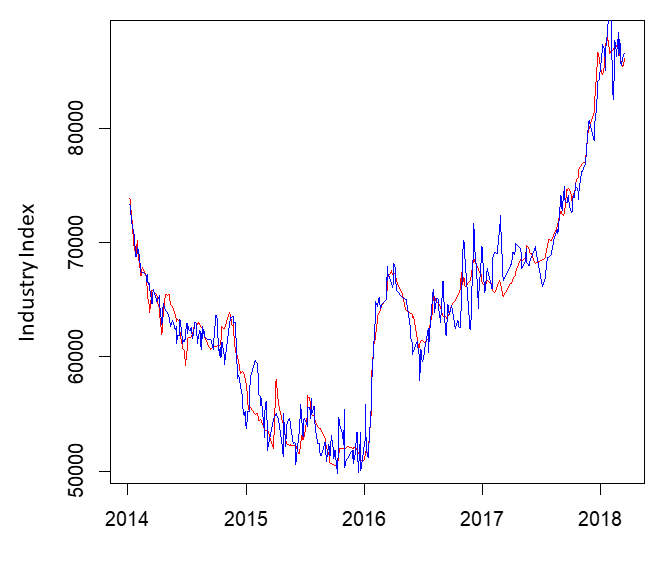}
\includegraphics[width=0.5 \textwidth]{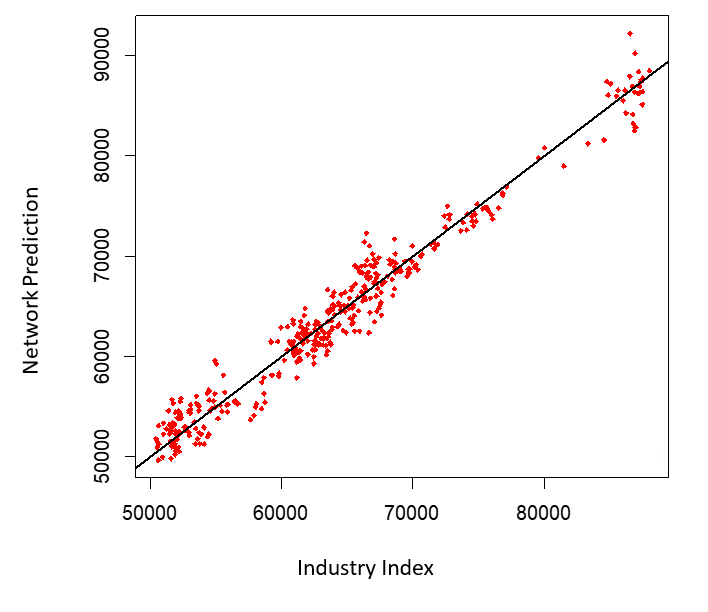}

\caption{The left figure shows Industry index prediction and actual data; The red line is actual values and blue line is feed-forward network prediction; The right figure represents Industry index and fitted line (perfect fit) In post-sanction and global financial crisis}
\label{fig:industry_post}       
\end{figure}

The second dataset starts from 2014 to 2018. In this period, international energy sanction was eased on Iran and oil price dropped drastically because of the global financial crisis. In this period, the network has 5 inputs (oil, gas and gold price, exchange rate, trading volume) and TEPIX and Industry index as outputs separately.  the number of computed nodes in hidden layers is 37. The learning continues till the network became converged. The results with 90 percent accuracy for both TEPIX and industry indices is presented in figure \ref{fig:stock_post} and \ref{fig:industry_post} respectively. Figures show that the feed-forward network has the ability to produce a good prediction by considering a wide range of economic variables. In this period, the model shows a better fit for Industry index in compare with the period of international energy sanction. 

The average percentage error (MPE), estimated root mean square error (RMSE) and mean absolute percentage error (MAPE) from learning FFNN for both TEPIX and industry index are listed in Table 2. 
Given the acceptable performance of FFNN in terms of accuracy, we can conclude that in the first period of study, international energy sanctions tighten on Iran, the model has a better performance for TEPIX (Table \ref{tab:criteria} shows the smaller error in compare with industry index). During this period, oil price was approximately steady. But industry index that is highly dependent on oil and gas companies, is influenced by the imposed sanctions on foreign investments and oil export. In 2014 to 2018, the model is a better fit for industry index that shows industry index is influenced by oil price shock.
During 2009 and 2014, FFNN performs the best in predicting stock index in compare with post international sanctions, which indicates the positive impact of oil price on the stock market in this period.   

%
\begin{table}[htbp]
\caption{Corresponding values of the evaluation criteria}
\label{tab:criteria}       
\begin{tabular}{lllll}
\hline\noalign{\smallskip}
& \multicolumn{2}{c}{International energy sanction}&\multicolumn{2}{c}{Post-sanction}\\
 Dependent variables&Stock index&Industry index &Stock index&Industry index\\
\noalign{\smallskip}\hline\noalign{\smallskip}
MAE & 0.107 & 0.116&0.107&0.09 \\
RMSE & 1106 & 1629&1750&1734 \\
MAPE & 0.07 & 0.16&0.23&0.33 \\

\noalign{\smallskip}\hline
\end{tabular}
\end{table}

\section{Conclusion}

In this paper, we model the impact of oil price volatility on Tehran stock and industry indices. 
To have a more realistic model, we consider a wide range of economic variables such as gas and gold price, exchange rate (Rial) and trading volume as explanatory variables.
We also analyze the aim of the study in two periods of international sanction and post-sanction to provide an overall picture on the impact of both sanctions and oil price shocks on an oil-dependent country such as Iran. We choose Iran as an ideal case for this setting because Iran is one of the largest oil exporters and has been under comprehensive sanctions. The result of feed-forward neural network with 90 percent accuracy indicates the positive impact of oil price on stock and industry indices, which is supported by empirical studies \cite{Fang2018} and \cite{Ewing2016} and \cite{Mezghani2018}. More specifically, the feed-forward neural networks performs better in predicting TEPIX in the period of international sanctions. 

In post-sanction and global financial crisis, the model evaluation criteria show a better value for industry index, which means Industry index is influenced by oil price shocks, as expected, industry index movements are more affected by oil price changes \cite{fang2017}.  
We can draw a conclusion the dependency of industry  on energy companies makes industry index more vulnerable to endogenous changes in oil market such as oil price shocks. 
In this paper, we have addressed the important question how changes in the international markets and politics influence stock market and industry in Iran.
In future research, we plan to explore the effect of uncertainty in international politics and markets on companies listed in the Tehran stock exchange.  
Following the impact of uncertainty in politics and international markets on Iran, future works may also pay attention to the changes in other important economic factors such as unemployment rate in Iran as more than 60 percent of Iran population are young professionals. An empirical extension of this paper is also comparing the results of FFNN with other non-linear models. 

\section*{Conflict of interest}
The authors declare that they have no conflict of interest.

\end{document}